\documentclass[aps,pre,reprint,groupedaddress]{revtex4-2}

\usepackage{bm}

\usepackage[colorlinks,linkcolor=blue,citecolor=magenta,urlcolor=blue]{hyperref}
\usepackage{amssymb}

\usepackage{latexsym, amsmath, mathrsfs, graphicx}
\usepackage[utf8]{inputenc}
\newcommand{\beq}{\begin{eqnarray}}
\newcommand{\eeq}{\end{eqnarray}}
\usepackage{xcolor}
\usepackage{esint}
\usepackage{empheq}
\usepackage[normalem]{ulem}
\usepackage[scr=boondox]{mathalfa}

\newcommand{\dd}{{\rm d}}

\newcommand{\Partial}[3]{\left( \frac{\partial #1}{\partial #2} \right)_{#3}}

\begin{document}

\title{Crossover from generalized to conventional hydrodynamics in nearly integrable systems under relaxation time approximation}

\author{Saikat Santra}
\email{saikat.santra@fuw.edu.pl}

\author{Maciej {\L}ebek}
\email{maciej.lebek@fuw.edu.pl}

\author{Mi{\l}osz Panfil}
\affiliation{Faculty of Physics, University of Warsaw, Pasteura 5, 02-093 Warsaw, Poland}

\date{\today}

\begin{abstract}
Upon breaking the integrability, the equations of generalized hydrodynamics (GHD) are supplemented by a Boltzmann collision term. Such terms are typically complicated and stem from a perturbative treatment of integrability-breaking terms in the hamiltonian. In our work, we study a simplified version of the collision operator in a form of relaxation time approximation familiar from kinetic theory. We explicitly compute transport coefficients which characterize the Navier-Stokes (NS) hydrodynamic regime emerging at large space-time scales. We also thoroughly study the crossover between GHD and NS hydrodynamic descriptions, identifying relevant characteristic space-time scales for the transition. In particular, we show how the emergence of NS hydrodynamics is visible in dynamics of conserved and non-conserved charge densities, and in hydrodynamic two-point functions.
\end{abstract}

\maketitle

\section{Introduction}
 Understanding the out-of-equilibrium behavior of interacting many-body quantum systems remains one of the most profound and active areas of research in theoretical physics. Generic isolated systems are expected to thermalize under non-equilibrium dynamics~\cite{Sachdev_PRA_2004,Thermalization_PRL_2004,Thermalization_nature_2008,Abanin2021}. However, certain class of complex systems have been identified that appear to avoid thermalization even after long times due to their integrable structure. To describe the hydrodynamics of such systems, a theoretical framework known as generalized hydrodynamics (GHD) has been developed~\cite{Bertini_PRL_2016,Doyon_PRX_2016,Doyon_PRL_2018,Doyon_scipost_2020,Doyon_JSM_2022,Doyon_PRX_2025}. It captures the essential characteristics of integrable dynamics, namely the existence of infinitely many conservation laws and the presence of stable, ballistically propagating quasiparticles. This framework has been successful in explaining various phenomena, such as the lack of thermalization observed in the well-known Newton's cradle experiment~\cite{Cradle_nature_2006,Cradle_PRX_2018}, the derivation of exact expressions for Drude weights~\cite{Doyon_scipost_2017,Jacopo_PRB_2017,Jacopo_PRL_2017,Bulchandani_PRB_2018} and the accurate description of anomalous transport in strongly interacting spin chains~\cite{Prosen_nature_2017,Varma_PRB_2018,Prosen_PRL_2018,Sarang_PRL_2019,Prosen_PRL_2019,Jacopo_PRL_2019}, to name a few. Recent developments~\cite{Cradle_nature_2006,Cradle_PRX_2018,Atomchip_PRL_2019,Wilson_science_2020,Li_scipost_2020,Moller_PRL_2021,Blocking_PRX_2022,Li_PRA_2023} in cold atomic systems, particularly in controlling and manipulating the interactions have fuelled the progress of GHD which by now has undergone several numerical and experimental verifications~\cite{Konik_PRL_2017,Spohn_JSM_2017,Nardis_PRR_2024,Rigol_science_2021, horvath2025,zeng2026,Dubois2026,Dubois2024,charnay2026}.
 
Integrable systems are fine-tuned and not exactly realized in experiments. Therefore, it is important to study nearly integrable systems where the integrability is weakly broken due to, for instance, additional interactions present among the particles. In such scenario a system is described by a Hamiltonian consisting of an integrable part, $\hat{H}_0$ and a weak non-integrable perturbation $\lambda \hat{V}$, where $\lambda$ is a small parameter controlling the strength of the integrability breaking. If the perturbation $\hat{V}$ describes an additional, long-range interactions between the particles and thus destroys all but three of the conserved quantities of the underlying integrable model, it has been shown that such systems eventually thermalize at times  $t \approx \mathcal{O}[(a\lambda)^{-2}]$~\cite{gBBGKY} where $a$ is the parameter controlling strength of integrable interactions. At large space-time scales, the effect of integrability breaking can be incorporated as corrections to the original GHD of the underlying integrable model~\cite{Moller_PRL_2021,Blocking_PRX_2022,Caux_scipost_2019,Friedman_PRB_2020,Dubail_scipost_2020,Luca_PRB_2020,Durnin_PRL_2021,GRTA_PRB_2021,Vasseur_JPA_2022,Vasseur_PRL_2023,Panfil_PRL_2023,Lebek2024}. Specifically, in the weak integrability breaking regime, the quasiparticle distribution $\rho_{\rm p}(\lambda,x,t)$ evolves according to the GHD equation modified with a collision integral $\mathcal{I}[\rho_{\rm p}]$, given by~\cite{Friedman_PRB_2020,Durnin_PRL_2021}
\begin{equation} \label{GHD_full}
	\partial_t \rho_{\rm p} + \partial_x \left(v \rho_{\rm p} \right) = \frac{1}{2} \partial_x \left(\mathfrak{D}\partial_x \rho_{\rm p} \right)+\mathcal{I}[\rho_{\rm p}],
\end{equation}
where $v$ is the effective velocity and $\mathfrak{D}$ represents the diffusion kernel~\cite{Bertini_PRL_2016,Doyon_PRL_2018,Doyon_PRL_2018}. Evaluating the collision integral $\mathcal{I}[\rho_{\rm p}]$ involves, either applying Fermi’s Golden Rule to the matrix elements of the integrability-breaking term~\cite{Friedman_PRB_2020,Durnin_PRL_2021} or using the gBBGKY hierarchy~\cite{gBBGKY}.

In this work, rather than computing the collision integral directly from a given integrability breaking perturbation of the system's Hamiltonian, we adopt an approximate form under the relaxation time approximation (RTA), as introduced in the GHD context in~\cite{GRTA_PRB_2021}:
\begin{equation}\label{eq:collision_integral}
\mathcal{I}_{\rm RTA}[\rho_{\rm p}]=\frac{\rho^{\rm th}_{\rm p}-\rho_{\rm p}}{\tau}.
\end{equation}
The fundamental assumption of RTA is that the quasiparticle distribution relaxes over a single characteristic timescale $\tau$. 
Remarkably, this assumption has been found to work well in reproducing the numerical time evolution in nearly-integrable XXZ model~\cite{GRTA_PRB_2021}.

Moreover, the RTA preserves the standard (conventional) conservation laws, making it particularly well-suited for studying hydrodynamic behavior. Consequently, one can argue that the RTA is capable of capturing the essential features of the crossover from generalized to conventional hydrodynamics—one of the central phenomena explored in this work.

It is expected that at short timescales ($t \ll \tau$), the effect of the collision integral on the dynamics is negligible and the system evolves according to GHD. In contrast, at long timescales ($t \gg \tau$), the dynamics is completely described by conventional hydrodynamics where only the conserved quantities respected by the collision integral play a role. If the collision integral destroys all conserved quantities except for the particle number, momentum and energy, the long-time dynamics is governed by the universal Navier-Stokes (NS) equations which are nothing but the continuity equations for the density, momentum and the energy fields~\cite{Balescu_book,Resibois_book,Dorfman_book,Lebek_PRL_2025,chapmanenskog}. 

In this work we present a detailed analysis of the crossover between the GHD and the conventional hydrodynamics for nearly integrable quantum gases within the RTA. We start with evaluating the transport coefficients for the Navier-Stokes equations. This amounts to solving certain integral equations that depend on the collision integral. For the RTA collision integral, as we show, they admit analytical solutions and are expressed in terms of the Drude weights and other thermodynamic quantities computed for integrable theory. This is discussed in Section~\ref{sec:transport_coefficients} and provides us with precise predictions of the conventional hydrodynamics. In the following Section~\ref{sec:cross_over}, we consider the full GHD dynamics and show that the conventional hydrodynamics emerges at late times and identify relevant space and time scales for the crossover. Here, we focus on a linear regime where initially homogeneous system is perturbed locally in space. This gives us an analytical control on the dynamics from which the emergence of the conventional hydrodynamics can be derived and understood. This allows us also
 to identify the relevant degrees of freedom which separate into the three standard conservation laws and the decaying modes. In the final Section~\ref{sec:correlations}, we apply our framework to study of the hydrodynamic correlation functions, in which a similar crossover between and GHD and conventional hydrodynamics is identified. Some of the technical detail of our work are to be found in the Appendices.

\section{Preliminaries}
We provide here a short overview of GHD and introduce ingredients needed in the later parts. Nearly integrable systems under the RTA can be described by GHD-Boltzmann equation of the form~\cite{Friedman_PRB_2020,Durnin_PRL_2021,Panfil_PRL_2023}
\begin{equation} \label{GHD_full_2}
	\partial_t \rho_{\rm p} + \partial_x \left(v_\rho \rho_{\rm p} \right) =\frac{1}{2} \partial_x \left(\mathfrak{D}_{\rho}\partial_x \rho_{\rm p} \right)+\mathcal{I}_{\rm RTA}[\rho_{\rm p}].
\end{equation}
The subscript $\rho$ in the effective velocity $v_{\rho}$ and the diffusion kernel $\mathfrak{D}_{\rho}$ indicates that both are functionals of the quasiparticle distribution $\rho_{\rm p}(\lambda,x,t)$ and therefore depend on space and time. However, for notational simplicity, we have omitted the subscript $\rho$ in \eqref{GHD_full} and will continue to do so hereafter. We consider Galilean-invariant models of particles with unit mass. While in this work we focus on particles with fermionic statistics having in mind the Lieb-Liniger model,  all the results can be easily generalized to arbitrary statistics by the means of changing statistical factor defined below and adjusting the expression for the free energy in terms of $\rho_{\rm p}$. For such generalization, we refer to~\cite{Doyon_scipost_2020}. In what follows we introduce the necessary ingredients of Thermodynamic Bethe Ansatz (TBA) and GHD, which are needed to understand the structure of GHD-Boltzmann equation.

Setting the right-hand side of \eqref{GHD_full_2} to zero corresponds to Euler GHD which is time-reversible and describes purely ballistic transport. For a quasiparticle with rapidity $\lambda$, the effective velocity $v(\lambda,x,t)$ refers to its group velocity, renormalized due to interactions with other quasiparticles inside the local state at position $x$ and time $t$. It can be obtained by solving the following linear integral equation~\cite{Doyon_PRX_2016,Doyon_scipost_2020}:
\beq
v(\lambda)=\lambda+2\pi \int \dd \lambda' \mathcal{T}(\lambda-\lambda') \rho_{\rm p}(\lambda') \left[ v(\lambda')-v(\lambda)\right ].
\label{eq:v_eff}
\eeq
The two-body scattering kernel  $\mathcal{T}(\lambda-\lambda')$ in \eqref{eq:v_eff} encodes the scattering shift due to a collision between two quasiparticles with rapidity $\lambda$ and $\lambda'$. The function $\mathcal{T}(\lambda)$ explicitly depends on the underlying microscopic interactions and is therefore model-dependent. In the case of the Lieb-Liniger model, $\mathcal{T}$ takes a particularly simple form,
\beq
\mathcal{T}(\lambda)=\frac{1}{2\pi} \frac{2c}{c^2+\lambda^2},
\label{eq:scattering_kernel}
\eeq
where $c$ denotes the strength of the $\delta$-function interaction potential~\cite{Lieb_1963,Yang1969}.

In addition to the Euler-scale GHD, a diffusion term appears in
the large scale dynamics of interacting integrable systems~\cite{Doyon_PRL_2018,Khemani_PRB_2018,Doyon_scipost_2019}. In such systems, diffusion originates from scatterings among quasiparticles, a feature absent in non-interacting integrable systems~\cite{Spohn_JMP_2018}. This diffusion term provides subleading corrections to the Euler-scale GHD and is responsible for positive entropy production as well as diffusive relaxation mechanisms~\cite{Doyon_PRL_2018,
Panfil_2019}. The diffusion operator in~\eqref{GHD_full_2} acts as $\partial_x \left[\mathfrak{D}\partial_x \rho_{\rm p} \right](\theta)=\int \dd \alpha \partial_x \left[\mathfrak{D}(\theta,\alpha) \partial_x \rho_{\rm p}(\alpha) \right]$ and
the diffusion kernel $\mathfrak{D}$ is given by
\begin{equation}\label{eq:diff_kernel}
\mathfrak{D}=(1-\mathscr{n}\mathcal{T})^{-1} \rho^{-1}_{\rm tot}\tilde{\mathfrak{D}} \rho^{-1}_{\rm tot} (1-\mathscr{n} \mathcal{T}),
\end{equation}
with $\tilde{\mathfrak{D}}(\lambda, \mu)=\delta (\lambda -\mu) w(\lambda)-W(\lambda, \mu)$. In our notation, $\mathscr{n}(\lambda)$ denotes the occupation function which is related to the rapidity distribution $\rho_{\rm p}(\lambda)$ through $\rho_{\rm p}(\lambda)=\mathscr{n}(\lambda)\rho_{\rm tot}(\lambda)$,
where $\rho_{\rm tot}$ is the total density of states. It satisfies the following integral equation~\cite{Doyon_scipost_2020}
\beq
\rho_{\rm tot}(\lambda)=\frac{1}{2\pi}+\int \dd \lambda' \mathcal{T}(\lambda-\lambda') \mathscr{n}(\lambda') \rho_{\rm tot}(\lambda').
\label{eq:density_dressing} 
\eeq
The two auxiliary functions appearing in  $\tilde{\mathfrak{D}}$ are
\begin{align}
W(\lambda, \mu)=\rho_{\rm p}(\lambda)f(\lambda) [\mathcal{T}^{\rm dr}(\lambda, \mu)]^2 |v(\lambda)-v(\mu)|,
\end{align}
and $w(\mu)=\int \dd \lambda W(\lambda, \mu)$.
The dressing operation applied to any function $g(\lambda)$ is defined as
\begin{align}
g^{\rm dr}(\lambda)= g(\lambda)+\int \dd \lambda' \mathcal{T} (\lambda-\lambda') \mathscr{n}(\lambda') g^{\rm dr}(\lambda').
\label{eq:dressing_operation}
\end{align}
At this point, let us comment on the recent refinements~\cite{Hubner2025} in the theory of GHD at the diffusive scale. The picture of a single equation with diffusion kernel~\eqref{eq:diff_kernel} is not correct in the presence of long-range correlations~\cite{Doyon2023LR}. Such correlations are dynamically generated in inhomogenous interacting integrable systems and one has to solve a system of two equations (for one- and two-point function) in order to properly describe the dynamics. In our case, we will always work in the linear regime, where the single equation with diffusion kernel $\mathfrak{D}$ is valid~\cite{Doyon2023LR}.

Lastly, we discus the Boltzmann kernel in RTA. The specific form of the collision integral~\eqref{eq:collision_integral} ensures that the system locally thermalizes to the Gibbs state over relaxation timescale $\tau$.  The state $\rho^{\rm th}_{\rm p}$ in~\eqref{eq:collision_integral} is a nonlinear functional of $\rho_{\rm p}$, determined by requiring that its conserved charges 
\begin{equation}\label{eq:compute_charges}
\mathscr{q}^{\rm th}_n(x,t)=\int \dd \lambda \mathscr{h}_n(\lambda)\rho^{\rm th}_{\rm p}[\rho_{\rm p}(\lambda,x,t)], \, \, n = 0, 1, 2,
\end{equation}
locally coincide with those of the state $\rho_{\rm p}$. By 
\begin{equation}
    \mathscr{h}_n(\lambda) =  \lambda^n/n!, \qquad n \in \mathbb{N},
\end{equation}
we denote single particle eigenvalues of the {\em ultralocal charges}.
In order to define thermal states we introduce more objects of TBA. In general, pseudoenergy $\epsilon(\lambda)$ fulfills the following integral equation
\begin{equation}
    \epsilon(\lambda) = \epsilon_0(\lambda) - \int {\rm d}\lambda' \mathcal{T}(\lambda - \lambda')\log\left(1 + e^{-\epsilon(\lambda')}\right),
\label{eq:epsilon_dressed}    
\end{equation}
where the bare pseudoenergy is given by
\begin{equation}
    \epsilon_0(\lambda) = \sum_{n} \beta_n \mathscr{h}_n(\lambda).
\end{equation}
A thermal boosted state is characterized by the first three chemical potentials, $\beta_0$, $\beta_1$ and $\beta_2$ while the remaining ($n \geq 3$) are zero.
From the pseudoenergy we can get the occupation function as $\mathscr{n}(\lambda) = 1/(1 + e^{\epsilon(\lambda)})$, which is then used to solve the integral equation~\eqref{eq:density_dressing} for $\rho_{\rm tot}$. 
Once $\rho^{\rm th}_{\rm p}(\lambda)$ is known, the expectation values of the conserved charges $\mathscr{q}^{\rm th}_n(x,t)$ can be computed using~\eqref{eq:compute_charges}. For a state defined by the distribution $\rho_{\rm p}$, the corresponding local thermal state is specified by the chemical potentials $\beta_0,\beta_1$ and $\beta_2$ which are adjusted so that the charges $\mathscr{q}^{\rm th}_n(x,t)$ coincide with $\mathscr{q}_n(x,t)$ obtained from $\rho_{\rm p}$. 

How the variation in chemical potentials changes the charges is encoded in the charge-susceptibility matrix $\mathcal{C}$, with kernel given as~\cite{Doyon_scipost_2020,Doyon_scipost_2017}
\beq
\mathcal{C}=(1-\mathscr{n}\mathcal{T})^{-1} \rho_{\rm p} f (1-\mathscr{n}\mathcal{T}).
\label{eq:matrix_susceptibility}
\eeq
Its matrix elements, in the basis of the ultralocal charges, are given by
\beq
\mathcal{C}_{mn}=-\frac{\delta \mathscr{q}_m}{\delta \beta_n} = \int \dd \lambda \rho_{\rm p}(\lambda) f(\lambda) \mathscr{h}^{\rm dr}_m(\lambda) \mathscr{h}^{\rm dr}_n(\lambda).
\label{eq:matrix_susceptibility_element0}
\eeq
We will also use the bra–ket notation for the matrix elements which makes the basis apparent,
\begin{equation}
    \mathcal{C}_{mn} = \langle \mathscr{h}_m | \mathcal{C} | \mathscr{h}_n \rangle.
\end{equation}
This is useful for introducing the hydrodynamic scalar product  $  \left( f | g \right) \equiv \langle f | \mathcal{C} | g \rangle$ and the special basis $\{h_n\}$ of {\em orthonormal charges} which diagonalize the hydrodynamic scalar product $(h_m|h_n) = \delta_{mn}$. They are constructed via the Gram-Schmidt (GS) orthonormalization of the ultralocal charges $\{\mathscr{h}_n(\lambda)\}$.
Similar to the charge-susceptibility matrix $\mathcal{C}$, one can define an analogous matrix $\mathcal{B}$ that characterizes the response of the associated currents $\{\mathscr{j}_m\}$ to changes in the chemical potentials $\{\beta_n\}$. The matrix $\mathcal{B}$ is given by
\beq
\mathcal{B}=(1-\mathscr{n}\mathcal{T})^{-1} \rho_{\rm p} f v (1-\mathscr{n}\mathcal{T}),
\label{eq:mathcalB_matrix}
\eeq 
with elements in the basis of ultralocal charges
\beq
\mathcal{B}_{mn}=-\frac{\delta \mathscr{j}_m}{\delta \beta_n}=\int \dd \lambda \rho_{\rm p}(\lambda) f(\lambda)v(\lambda) \mathscr{h}^{\rm dr}_m(\lambda) \mathscr{h}^{\rm dr}_n(\lambda).
\label{eq:matrix_susceptibility_element2}
\eeq
Finally, we are interested in the linearized GHD dynamics where $\rho_{\rm p} = \rho_{\rm p}^{\rm th} + \delta \rho_{\rm p}$. Here $\rho_{\rm p}^{\rm th}$ is a homogeneous thermal state and $\delta \rho_{\rm p}$ is a small perturbation of quasiparticle density on top of it. Linearizing Eq.~\eqref{GHD_full} about the thermal state $\rho^{\rm th}_{\rm p}$, one gets
\begin{align}
\partial_t \delta \rho_{\rm p}
+ \mathcal{A}_{\rm th}\partial_x
\delta \rho_{\rm p}  = \frac{1}{2}\mathfrak{D}_{\rm th} \partial^2_x   \delta \rho_{\rm p} 
+ \left( \frac{\delta \mathcal{I}}{\delta \rho_{\rm p}} \right)_{\rm th}\delta \rho_{\rm p},
\label{eq:GHD_linear}
\end{align}
where the operator $\mathcal{A}=(1-\mathscr{n}\mathcal{T})^{-1} v_{\rho} (1-\mathscr{n} \mathcal{T})$~\cite{Doyon_scipost_2017} and the subscript `th' indicates that these quantities are evaluated in the homogenous thermal state $\rho_{\rm p}^{\rm th}$.
It is convenient to express the resulting dynamics in terms of perturbations to the orthonormal charges, $ \delta q_n(x,t) = \int {\rm d}\lambda h_n(\lambda) \delta \rho_{\rm p}(\lambda, x,t)$.
The equations are~\cite{Lebek_PRL_2025}
\begin{equation}
\partial_t \delta q_n+\sum_{m}A_{nm} \partial_x \delta q_m= \sum_{m} \mathcal{D}_{nm} \partial^2_x \delta q_m -\sum_{m} \Gamma_{nm} \delta q_m.
\label{eq:charge_evolution}
\end{equation}
The matrices $A$, $\mathcal{D}$ and $\Gamma$ have the following elements,
\begin{align}
A_{nm}=& \int \dd \lambda \rho_{\rm p}(\lambda) f(\lambda) v(\lambda) h^{\rm dr}_n(\lambda)   h^{\rm dr}_m(\lambda),  \\
 \mathcal{D}_{nm}=&\frac{1}{2} \int \dd \lambda \dd \mu h^{\rm dr}_n(\lambda) \rho^{-1}_{\rm tot}(\lambda) \tilde{\mathfrak{D}}_{\rm th}(\lambda,\mu), \\
 &~~~~~~~~~~~~~~~~~~~~~\times \rho^{-1}_{\rm tot}(\mu) \rho_{\rm p}(\mu) f(\mu) h^{\rm dr}_m(\mu) \\
\Gamma_{nm}=&\int \dd\lambda \dd\mu h_n(\lambda) \Gamma(\lambda,\mu) h_m(\mu),
\label{eq:ADT_matrix}
\end{align}
where the collisional operator $\Gamma$ is obtained from the functional derivative of the collision integral with respect to the quasiparticle density and takes the form
\beq
\Gamma=-\frac{\delta \mathcal{I}}{\delta \rho_{\rm p}} \mathcal{C}.
\label{eq:gamma_operator}
\eeq 
As a last ingredient we introduce Drude weights which measures the ballistic, dissipationless part of transport in a system~\cite{Doyon_scipost_2017}. The matrix elements are given by (note that these are matrix elements in orthonormal basis)
\begin{equation} \label{drude_weights}
    D_{mn}= \int \dd \lambda \rho_{\rm p}(\lambda) f(\lambda) v^2(\lambda)h_m^{\rm dr}(\lambda) h_n^{\rm dr}(\lambda). 
\end{equation}
We now turn to the analysis of transport coefficients in nearly integrable systems within the RTA.

\section{Navier-Stokes equations and transport coefficients} \label{sec:transport_coefficients}
The first three conserved charges $(n=0,1,2)$ in the Galilean invariant model correspond to particle number, momentum and energy. If the collision integral respects these three conservation laws i.e., 
\begin{equation}
\int \dd \lambda \mathscr{h}_n(\lambda) \mathcal{I}(\rho_{\rm p})=0,\qquad n=0,1,2,
\end{equation}
and breaks all the other then at large space-time scales the dynamics can be described by the Navier-Stokes equations~\cite{Lebek_PRL_2025,chapmanenskog,Balescu_book,Resibois_book,Dorfman_book},
\begin{equation} 
\begin{aligned}
		\partial_t \varrho &= - \partial_x (\varrho u), \quad \partial_t (\varrho u) = - \partial_x (\varrho u^2 + \mathcal{P}), \\
		\partial_t (\varrho e) &= - \partial_x (u \varrho e+\mathcal{J}) - \mathcal{P} \partial_x u.
\end{aligned}
\label{eq:NS_equations}
\end{equation}
The hydrodynamic fields are related to the local densities through the following relations~\cite{Lebek_PRL_2025}:
\begin{equation}
	\varrho = \mathscr{q}_0, \qquad u = \frac{\mathscr{q}_1}{\mathscr{q}_0}, \qquad e = \frac{\mathscr{q}_2}{\mathscr{q}_0} - \frac{\mathscr{q}_1^2}{2 \mathscr{q}_0^2}.
\label{eq:fields_conserved}    
\end{equation}
The dynamic pressure $\mathcal{P}$ and heat current $\mathcal{J}$ are determined by the Newton's law for a viscous fluid (with shear viscosity absent in one spatial dimension) and the Fourier's law for the heat current:
\begin{equation}
	\mathcal{P} = P - \zeta \partial_x u, \qquad \mathcal{J} = - \kappa \partial_x T. \label{pressure_current}
\end{equation}
Here, $P$ denotes the standard thermodynamic pressure, $T$ the local temperature and the two transport coefficients are viscosity $\zeta$ and thermal conductivity $\kappa$. They receive additive contributions which can be identified as one coming from the GHD diffusion, hence due to integrable interactions, and the other from the collision integral, hence due to integrability breaking interactions,
\beq
\zeta = \zeta_{\mathfrak{D}} + \zeta_{\mathcal{I}}, \qquad  \kappa = \kappa_{\mathfrak{D}} + \kappa_{\mathcal{I}}.
\label{eq:transport_coefficients}
\eeq

In Ref.~\cite{Lebek_PRL_2025}, the contributions $\zeta_\mathfrak{D}$ and $\kappa_\mathfrak{D}$, which arise from the GHD diffusion and are independent of the details of the collision integral, have been thoroughly investigated. They have the following forms
\beq
\zeta_{\mathfrak{D}}=\frac{1}{2T} \langle \mathscr{h}_1 | \mathfrak{D} \mathcal{C} |\mathscr{h}_1 \rangle, \qquad \kappa_{\mathfrak{D}} =  \frac{1}{2 T^2} \langle \mathscr{h}_2 | \mathfrak{D} \mathcal{C} |\mathscr{h}_2 \rangle.
\label{eq:transport_coeff_diff}
\eeq
On the other hand, the contributions $\kappa_\mathcal{I}$ and $\zeta_\mathcal{I}$ are due to the collision integral and depend on their precise structure. They capture dissipative effects of quasiparticle scattering and are given by
\beq
\zeta_{\mathcal{I}}=  \langle \mathscr{h}_{1}|\mathcal{B} |\phi_{\zeta} \rangle, \qquad  \kappa_\mathcal{I}=  \langle \mathscr{h}_{2}|\mathcal{B} |\phi_{\kappa} \rangle.
\label{eq:conductivity_integral_equation}
\eeq
The functions $\phi_{\zeta}$ and $\phi_{\kappa}$ are determined by integral equations involving the linearized collision operator
\beq
\Gamma | \phi_{\zeta,\kappa} \rangle = |\eta_{\zeta,\kappa} \rangle,
\label{eq:equation_Gamma_epsilon}
\eeq
together with the relations $\langle h_n |\mathcal{C}| \phi _{\zeta,\kappa} \rangle = 0$ for $n=0,1,2$. The source terms $\eta_{\zeta,\kappa}$ are given by~\cite{Lebek_PRL_2025}
\begin{align}
\eta_{\zeta}= &\sqrt{\frac{\varrho}{T}} \left[\vphantom{\left (h_0+\sqrt{\frac{c_P-c_V}{c_V}} h_2 \right )} \mathcal{B} h_1- \frac{1}{\sqrt{\varrho \varkappa_T}} \mathcal{C} \left (h_0+\sqrt{\frac{c_P-c_V}{c_V}} h_2 \right ) \right ], \\
\eta_{\kappa}=& \frac{\sqrt{\varrho c_V}}{T} \mathcal{B} \left (h_2-\sqrt{\frac{c_P-c_V}{c_V}} h_0 \right ). 
\label{eq:eta_functions}
\end{align}
Here, $\varkappa_T$ represents the isothermal compressibility of the system while the specific heats at constant volume and constant pressure are denoted by $c_V$ and $c_P$, respectively. All thermodynamic quantities are computed for the integrable model using TBA. The $\Gamma$ operator can be diagonalised (it is real and symmetric), $\Gamma |\gamma_\alpha\rangle = \gamma_\alpha |\gamma_\alpha\rangle$ and the solution to the equation~\eqref{eq:equation_Gamma_epsilon} is
\begin{equation}
    | \phi_{\zeta,\kappa} \rangle = \sum_{\gamma_\beta \neq 0} \gamma_{\beta}^{-1} \langle g_\beta | \eta_{\zeta, \kappa} \rangle | g_\beta\rangle. 
\end{equation}
As we shown in Appendix~\ref{app:gamma}, it is the orthonormal basis $\{h_n\}$ that diagonalizes the RTA $\Gamma$ operator. Consequently, it has three zero eigenvalues, corresponding to the three collision invariants, while all the other eigenvalues are equal $\tau^{-1}$, a feature specific to the RTA collision integral. This gives
\begin{equation}
    | \phi_{\zeta,\kappa} \rangle = \tau\sum_{n \neq 0,1,2} \langle h_n | \eta_{\zeta, \kappa} \rangle | h_n\rangle,
\end{equation}
with the relations $\langle h_n |\mathcal{C}| \phi_{\zeta,\kappa} \rangle = 0$ for $n=0,1,2$ fulfilled. The formulas for the transport coefficients are then 
\begin{align}
\zeta_\mathcal{I}&= \tau \sum_{n \neq 0,1,2} \langle \mathscr{h}_1 | \mathcal{B} | h_n \rangle \langle h_n | \eta_{\zeta} \rangle, \label{zetaI_simplified}\\
\kappa_\mathcal{I}&= \tau \sum_{n \neq 0,1,2} \langle \mathscr{h}_2 | \mathcal{B} | h_n \rangle \langle h_n | \eta_{\kappa} \rangle.
\label{eq:conductivity_expre_2}
\end{align}
\begin{figure*}[t]
\centering
\includegraphics[width=\textwidth]{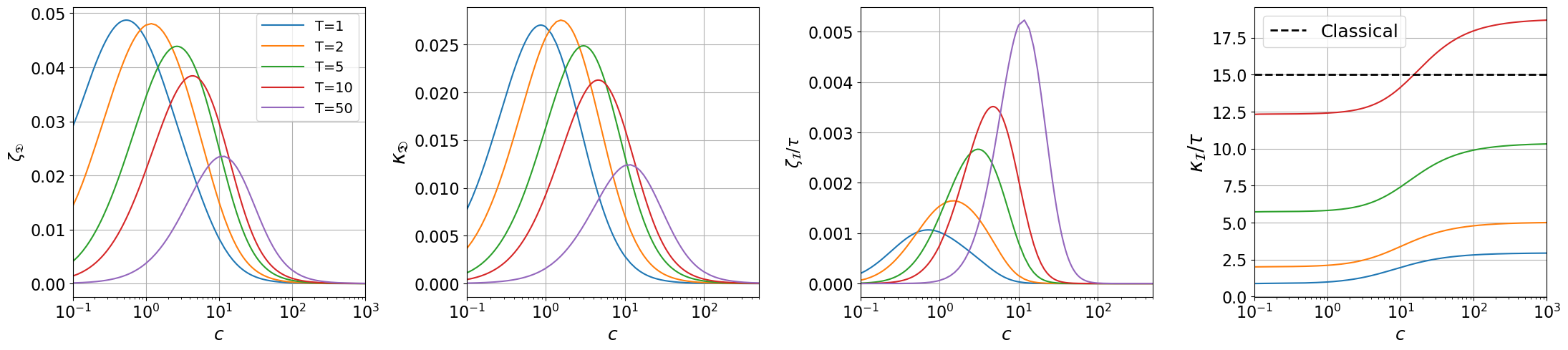}
\caption{Contributions to the transport coefficients as a function of the interaction strength $c$ for several values of temperature $T$. The contributions $\zeta_{\mathfrak{D}}$ and $\kappa_{\mathfrak{D}}$ are computed using \eqref{eq:transport_coeff_diff} while the contributions $\zeta_{\mathcal{I}}$ and $\kappa_{\mathcal{I}}$ are obtained from \eqref{eq:transport_coeff_coll}. Irrespective of the temperature $T$, all coefficients except  $\kappa_{\mathcal{I}}$ show non-monotonic behavior as a function of $c$. Starting near zero for small $c$, they attain a maximum value at a finite value of $c$ and eventually vanish at very large $c$. With increasing temperature $T$, the positions of these maxima shift toward larger values of $c$. In contrast, the $\kappa_{\mathcal{I}}$ displays monotonic behavior with respect to $c$, it begins at a finite value at $c \to 0$, increases monotonically with $c$ and asymptotically approaches  another finite value at $c \to \infty$. For reference, a dashed line corresponding to classical free particles, given by the formula $\kappa_{\mathcal{I}}=\frac{3}{2} \varrho \tau T$, which is derived from simple adaptation of \eqref{eq:omega1_simplified} is shown in the last panel for the temperature $T=10$. For better visibility, we refrain from plotting the data for $\kappa_{\mathcal{I}}$ at temperature $T=50$.  In all figures, the density $\varrho$ is set to $1$.}
\label{fig:transport_coefficients}
\end{figure*}
After straightforward algebraic manipulations, shown in Appendix~\ref{app:transport}, these expressions  can be simplified to
\beq
~\zeta_\mathcal{I}=\varrho \tau \omega^{\rm coll}_1, \qquad \kappa_\mathcal{I}=\varrho c_V \tau  \omega^{\rm coll}_{\rm th},
\label{eq:transport_coeff_coll}
\eeq
with
\begin{align}
\omega^{\rm coll}_{1} &=D_{11} - A_{10}^2-A_{12}^2, \quad \omega^{\rm coll}_{\rm th} &=  D_{22} - A_{12}^2.
\label{eq:omega1_simplified}    
\end{align}
Note that, the matrix elements $A_{10}$ and $A_{12}$ in~\eqref{eq:omega1_simplified}  are explicitly known in terms of the thermodynamic parameters of the system~\cite{Lebek_PRL_2025}:
\beq
A_{10}=\frac{1}{\sqrt{\varrho \varkappa_T}}, \quad A_{12}=\frac{1}{\sqrt{\varrho \varkappa_T}} \sqrt{\frac{c_P-c_V}{c_V}}.
\label{A_matrix_thermo}
\eeq
In Fig.~\ref{fig:transport_coefficients}, we plot the $\mathfrak{D}$ and $\mathcal{I}$ components of the thermal conductivity and the viscosity separately as a function of the interaction strength $c$. Apart from the $\kappa_{\mathcal{I}}$, all other contributions exhibit non-monotonic behaviour as a function of $c$: approaching zero in the limit $c \to 0$, they reach a maximum at some finite value of $c$ and then vanish at $c \to \infty$. The limiting values of the $\mathfrak{D}$ components are already known, since at the two special limits ($c \to 0$ and $c \to \infty$) the system becomes non-interacting and therefore there are no diffusive corrections to the GHD equation~\cite{Doyon_PRL_2018}. The fact that the coefficient $\zeta_{\mathcal{I}}$ also vanishes in these two limits is shown explicitly in Appendix.~\ref{sec:comp_limiting_w1}. Several curves are shown corresponding to different system temperature. In the strongly interacting regime i.e, when $c \gg 1$, the transport coefficient increases with increasing temperature. For smaller $c$, quantum effects become significant, and the temperature dependence is reversed. The $\kappa_{\mathcal{I}}$, on the other hand, behaves differently: it starts from a finite value at $c \to 0$, increases monotonically with $c$, and asymptotically approaches another value at $c \to \infty$. Irrespective of the value of $c$, its numerical value increases with increasing temperature.

In summary, we have obtained the expressions~\eqref{eq:transport_coeff_diff} and~\eqref{eq:transport_coeff_coll} for the transport coefficients $\zeta$ and $\kappa$ in terms of the known thermodynamic matrices such as $\tilde{\mathfrak{D}}, D$ and $A$. We have illustrated their behavior in Fig.~\ref{fig:transport_coefficients}. In the next section, we discuss the crossover from GHD to NS dynamics.

\section{GHD-NS crossover} \label{sec:cross_over}

In this section, we investigate how different quantities—both conserved and non-conserved evolve in a nearly integrable systems within the RTA framework. Specifically, we consider the following scenario: a small amount of charge, such as particles or energy, is deposited at a specific location and we observe how it spreads across the system over time.
Since transport phenomena are typically described by small deviations from local equilibrium, it is sufficient to consider linear perturbations around a homogeneous thermal background. This approximation allows for a systematic study of how local disturbances propagate and relax over time.

We begin by discussing the space-time evolution of the conserved charges as governed by the NS equations. Subsequently, we address both the conserved and non-conserved charges using the GHD-Boltzmann equation. 

To set up the stage, we start with recalling the basic facts about Navier-Stokes equations in the linear regime, which will serve us as a reference point in the study of GHD-Boltzmann dynamics.

\subsection{Navier-Stokes hydrodynamics in linear regime}
In what follows it will be convenient to change the degrees of freedom in NS equations~\eqref{eq:NS_equations} using the thermodynamic identity~\cite{resibois}
\beq
\delta T =\frac{1}{c_V} \delta e -\frac{1}{c_V \varrho^2}\left [P-T\left (\frac{\partial P}{\partial T}\right )_{\varrho} \right ]\delta \varrho.
\eeq
Our motivation here is to solve NS equations in the linear regime by considering perturbations around homogeneous thermal state.
The relevant equations are~\cite{Resibois_book}
\begin{align}
 \partial_t \delta \varrho & =-\varrho \partial_x \delta u, \notag \\
\partial_t \delta u & =-\frac{1}{\varrho} \left ( \frac{\partial P}{\partial \varrho} \right )_T \partial_x \delta \varrho -\frac{1}{\varrho} \left ( \frac{\partial P}{\partial T} \right )_{\varrho} \partial_x \delta T +\frac{\zeta}{\varrho} \partial^2_x \delta u, \notag \\
\partial_t \delta T & =-\frac{T}{\varrho c_V} \left (\frac{\partial P}{\partial T} \right )_{\varrho} \partial_x \delta u +\frac{\kappa}{\varrho c_V} \partial^2_x \delta T.
\label{eqs:NS_linearized}
\end{align}
They can be solved with Fourier-Laplace transform
\begin{equation}\label{eq:FLtransf}
    \delta \varrho(x,t) = \int {\rm d} k \, e^{ikx-\Lambda^{\rm NS}(k)t} \delta \tilde{\varrho} (k,0),
\end{equation}
and similarly for $\delta u, \delta T$, which turns them into $k$-dependent $3 \times 3$ eigenproblem.
To make the associated matrix symmetric, it is convenient to change the degrees of freedom to $\delta q_0, \, \delta q_1, \, \delta q_2$ defined as follows~\cite{Lebek_PRL_2025}
\begin{equation}
    \delta \varrho= \varrho \sqrt{T \varkappa_T} \delta q_0, \;\;\; \delta u = \sqrt{T}{\varrho} \delta q_1, \;\;\;  \delta T =\frac{T}{\sqrt{\varrho c_V}} \delta q_2.
\end{equation}
The eigenvalue problem then reads 
\begin{equation}
    \mathcal{M}^{\rm NS}_{nm}(k)\delta \tilde{q}_m(k) = \Lambda^{\rm NS}(k)\delta \tilde{q}_n(k) , \quad n=0,1,2,
\end{equation}
with summation over $m=0,1,2$ is implied and with $\mathcal{M}^{\rm NS}(k)$ given by
\begin{equation}
    \mathcal{M}^{\rm NS}(k) = \begin{pmatrix}
     0 & i \frac{1}{\sqrt{\varrho \varkappa_T}} k & 0 \\
    i  \frac{1}{\sqrt{\varrho \varkappa_T}} k &  \frac{\zeta}{\varrho} k^2 & i \sqrt{\frac{T}{\varrho^2 c_V}} \Partial{P}{T}{\varrho} k \\
    0 & i \sqrt{\frac{T}{\varrho^2 c_V}} \Partial{P}{T}{\varrho} k & \frac{\kappa}{\varrho c_V} k^2
    \end{pmatrix}.
\end{equation}
Let us denote by $R^{\rm NS}(k)$ the matrix which diagonalizes the problem: ($R^{\rm NS})^{-1}(k)\mathcal{M}^{\rm NS}(k) R^{\rm NS}(k) = \Lambda^{\rm NS}(k)$.
We find two sound modes and a heat mode, with dispersion relations for small $k$ given by
\begin{equation}\label{eq:NSdisp}
\begin{aligned}
\Lambda^{\rm NS}_{\pm}(k)=&\pm i v_s k +\frac{1}{2 \varrho} \left [ \left (\frac{1}{c_V}-\frac{1}{c_P} \right ) \kappa +\zeta  \right ] k^2  +\mathcal{O}(k^3), \\
\Lambda^{\rm NS}_{\rm th}(k)=&\frac{\kappa}{\varrho c_P} k^2 +\mathcal{O}(k^3),
\end{aligned}
\end{equation}
where $v_s$ denotes the speed of sound and can be derived from the thermodynamics of the system~\cite{Resibois_book,Lebek_PRL_2025}. The corresponding eigenstates (sound modes and heat mode) are encoded in the matrix $R^{\rm NS}(k)$ in $k \to 0$ limit:
\begin{equation}
    R^{\rm NS}(k) = \begin{pmatrix}
    \frac{1}{\sqrt{2}}\sqrt{\frac{c_V}{c_P}} & -1 & \sqrt{\frac{c_P-c_V}{c_P}} \\
    -\sqrt{\frac{c_P-c_V}{c_P}} & 0 & \sqrt{\frac{c_V}{c_P}} \\
     \frac{1}{\sqrt{2}}\sqrt{\frac{c_V}{c_P}} & 1 & \sqrt{\frac{c_P-c_V}{c_P}}
    \end{pmatrix}.
\end{equation}
Lastly, let us write the explicit dynamics of the fields $\delta \tilde{q}_n$. It is given by the following expression
\begin{equation}\label{eq:charge_dyn_NS}
    \delta \tilde{q}_n(k,t) =\sum_{\sigma,p}(R^{\rm NS})^{-1}_{n \sigma}(k) e^{-\Lambda^{\rm NS}_{\sigma}(k)t} R^{\rm NS}_{\sigma p}(k) \delta \tilde{q}_p(k,0),
\end{equation}
where the index $\sigma$ runs over the eigenmmodes $\sigma=+,-,{\rm th}$, while the indices $n$ and $p$ run over $0,1,2,$ labelling the charge perturbations.

\subsection{GHD-Boltzmann equation in linear regime}

We now turn to the linearized GHD-Boltzmann equation~\eqref{eq:charge_evolution}. It can be simplified using Fourier-Laplace transform, similarly to~\eqref{eq:FLtransf}. This transforms Eq.~\eqref{eq:charge_evolution} into $k$-dependent eigenvalue problem
\begin{equation}\label{eq:eigenproblem}
    \Lambda(k) \delta \tilde{q}_n(k)=\sum_m\mathcal{M}_{nm}(k) \delta \tilde{q}_m(k),
\end{equation}
where the matrix elements of $\mathcal{M}(k)$ are given by
\begin{equation}\label{eq:Mmatrix}
\mathcal{M}_{nm}(k)=\left(\Gamma_{nm}+i k A_{nm}  + k^2 \mathcal{D}_{nm}\right).
\end{equation}

\begin{figure}[t]
    \includegraphics[scale=0.4]{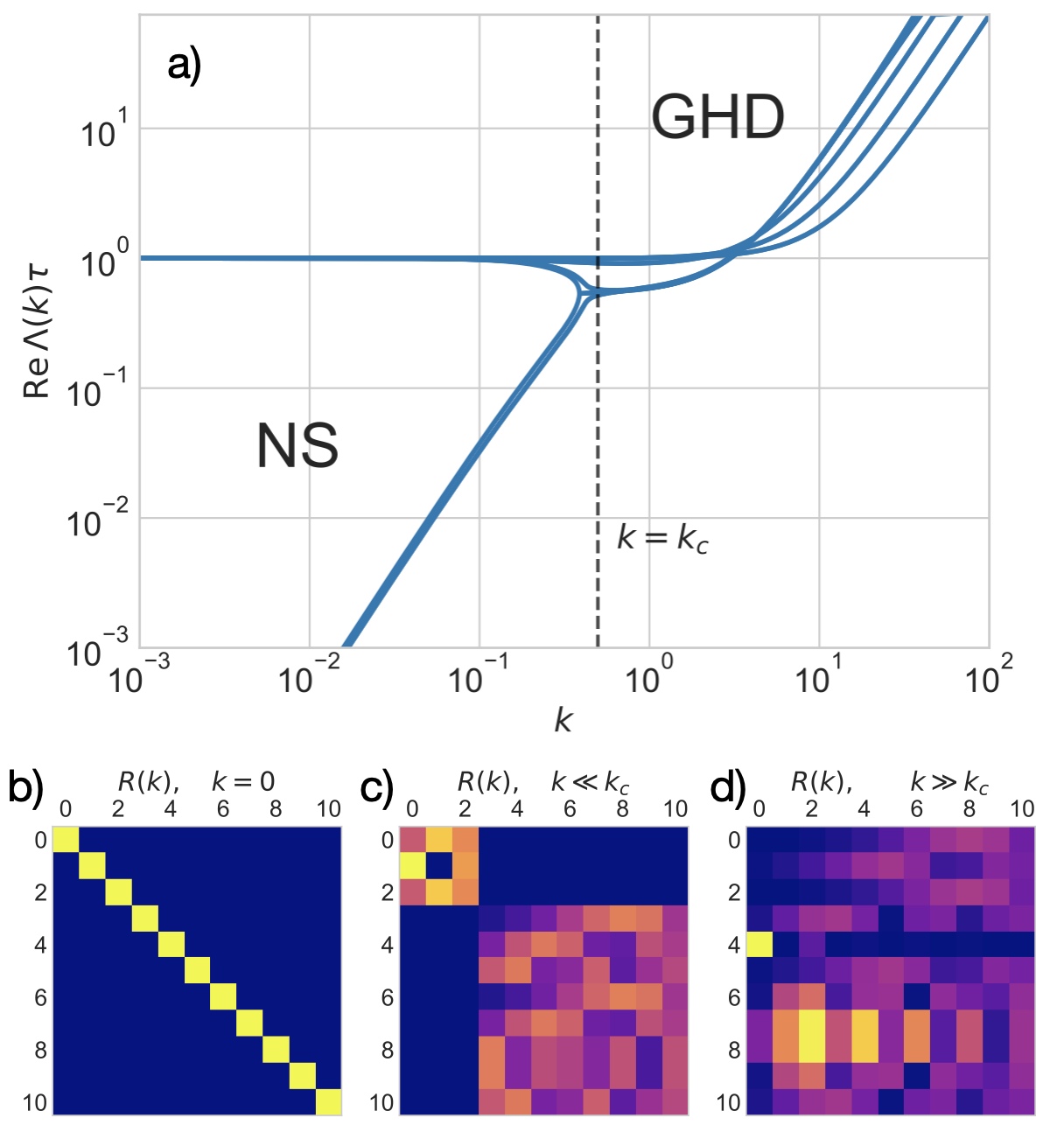}
    \caption{Panel (a): real part of the spectrum of matrix $\mathcal{M}(k)$ as a function of $k$. We consider interaction coupling $c=2$, temperature $T=4.36$ and density $\varrho=1$ with RTA timescale $\tau=1$. For small momenta $k \ll k_c$ we observe gapless modes (in fact there are three such modes, the sound modes have degenerate $\text{Re}\Lambda(k)$) and gapped ones with a gap $\tau^{-1}$. At larger momenta $k \gg k_c$ we find the modes of GHD. Panels (b)-(d): absolute value of rotation matrix $R(k)$ in three different regimes. For $k=0$ we have $\mathcal{M} = \Gamma$ and the operator is diagonal. For $k$ finite, but much smaller than $k_c$ we observe a NS sector for the first three charges. The coefficients are those of standard sound and heat modes. For $k\gg k_c$ the modes of the systems are GHD modes, which are not simply related to the basis $\{ h_n \}$. This is reflected in featureless $R$ matrix.}
    \label{fig:Spectrum_crossover}
\end{figure}
Let $R(k)$ denote the matrix which diagonalizes the matrix $\mathcal{M}(k)$ such that $R^{-1}(k) \mathcal{M}(k) R(k)=\mathbf{\Lambda}(k)$ with $\mathbf{\Lambda}(k) ={\rm diag} \left(\Lambda_1(k), \Lambda_2(k), \dots \right)$. The dynamics of charges, in complete analogy to~\eqref{eq:charge_dyn_NS}, is then given by
\begin{equation}\label{eq:charge_dyn}
\delta \tilde{q}_n(k,t)= \sum_{m,p} R_{nm}^{-1}(k) e^{-\Lambda_m(k) t}R_{mp}(k)\delta \tilde{q}_p(k,0),
\end{equation}
where $m,p=0,1,2,\dots$ run over all charges (both conserved or non-conserved). The solution~\eqref{eq:charge_dyn} represents the exact dynamics in linearized regime of GHD-Boltzmann equation. It involves the data from eigenvalue problem~\eqref{eq:eigenproblem} in the form of eigenvalues $\Lambda(k)$ together with the rotation matrix $R(k)$ and the initial state $\delta \tilde{q}_p(k,0)$. In what follows we will show how the properties of these objects can be used to understand and characterize the GHD-NS crossover, controlled by the (inverse) spatial scale $k$ and time $t$. Even though the picture found by us is simpler in RTA, the intuitions presented here should hold for more general form of collision integral $\mathcal{I}[\rho_{\rm p}]$. Indeed, all the qualitative features mentioned above were also found in the analysis for FGR collision integral describing coupled Lieb-Liniger models~\cite{Lebek_PRL_2025}.
 \begin{figure*}[t]
\centering
\includegraphics[width=\textwidth]{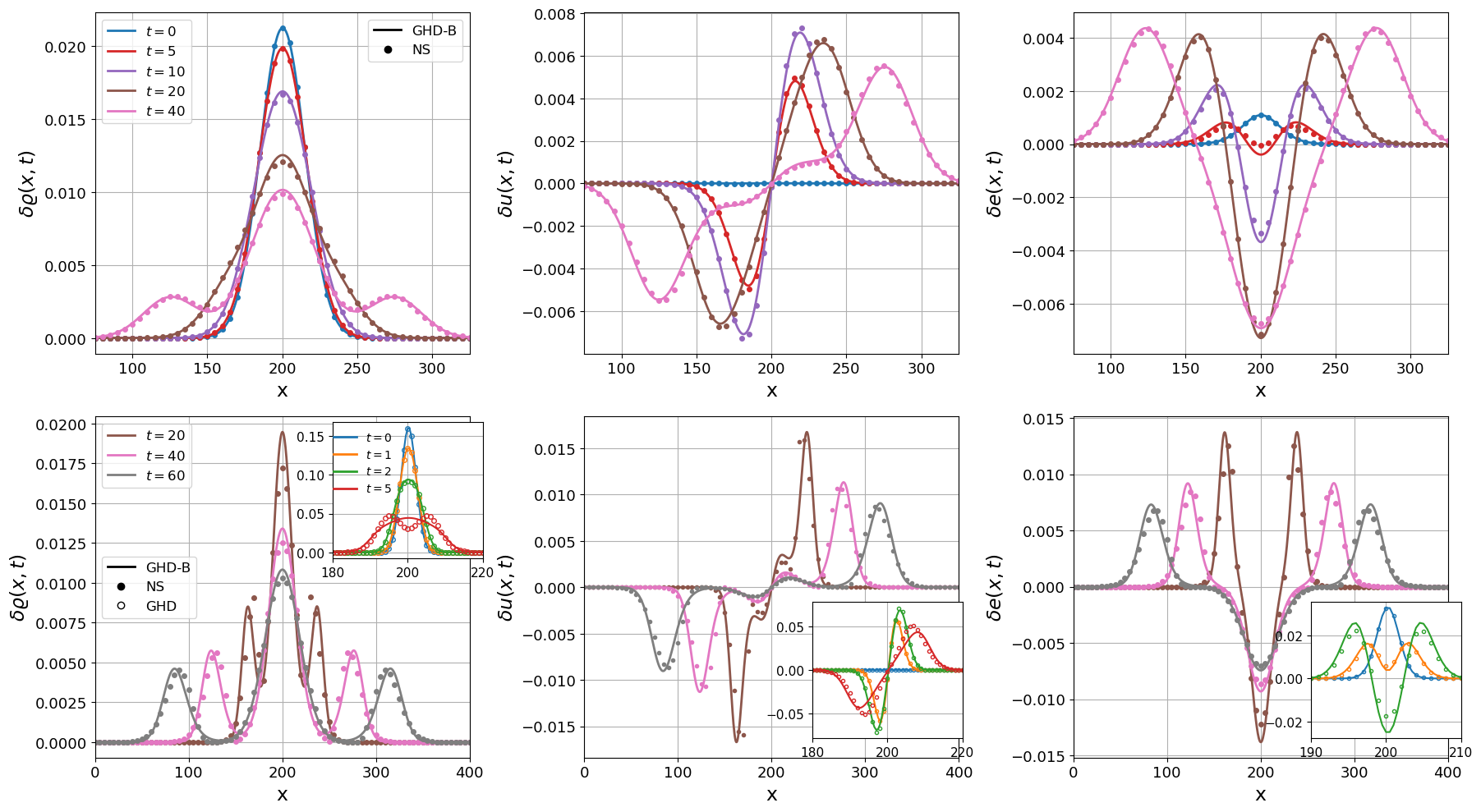}
\caption{Dynamics of the density, velocity and energy fields starting from an initial perturbation in the $\delta \varrho$ and $\delta e$ fields around a homogeneous thermal state $(\varrho^{\rm th}=\varrho, u^{\rm th}=0, T^{\rm th}=T)$. The perturbation is evolved using the linearized NS equations, \eqref{eqs:NS_linearized}, with the transport coefficients $\zeta$ and $\kappa$ computed from \eqref{eq:transport_coefficients}. These results are shown in the main plots as solid circles labeled ``NS” at different times during the evolution. They are then compared with the corresponding fields obtained by solving the linearized GHD–Boltzmann equation, \eqref{eq:charge_evolution}, shown as solid lines labeled ``GHD-B.”. The agreement between the two at late times confirms that the large-scale space–time behavior of the conserved fields is accurately captured by the NS dynamics with correctly computed transport coefficients. The parameters used here are: $c = 2.0$ and $\tau = 2.0$, $\varrho=1.0$, $T=1.36$. In this setup, the crossover momentum $k_c$ obtained from \eqref{eq:k_critical} is approximately $0.57$, while the characteristic momentum
 of initial perturbation is $k_{\rm ini} \approx0.25$ (upper panels) and $k_{\rm ini} \approx 1.9$ (lower panels). In the top row, where $k_{\rm ini}<k_c$, the original dynamics (GHD-Boltzmann) match NS results even at short times $t \sim \tau$. In contrast, in the bottom row, where $k_{\rm ini}>k_c$, it takes longer time $(t \sim 10 \tau )$ for the dynamics to converge to the corresponding NS fields. In the insets of the lower panels, we compare our GHD-Boltzmann data with the pure GHD results [i.e., without the collision term in \eqref{GHD_full}], shown as void circles labeled ``GHD”. At very short times, $t \sim \tau$, the two results match, indicating that the effect of collision integral has not yet become significant. As time increases further, deviations between the two start to appear and at very large time NS dynamics start to appear which is shown in the main plots.}
\label{fig:comparison_fields}
\end{figure*}

Firstly, it is shown in Appendix~\ref{app:gamma} that within the RTA,
the orthonormalized charge basis $\{h_n(\lambda)\}$ diagonalizes the matrix $\Gamma$. The matrix element $\Gamma_{nm}$ appearing in Eq.~\eqref{eq:ADT_matrix} simplifies to
\begin{equation}\label{eq:matrix_element_taunm}
\Gamma_{nm}= \delta_{nm} 
    \begin{cases}
        0,  &n = 0,1,2, \\
        \tau^{-1}, &n \neq 0,1,2.
    \end{cases}
\end{equation}
Let us now analyze the eigenvalues of~\eqref{eq:Mmatrix}. For generic $k$ it is not possible to find the spectrum analytically. However, for small $k$ the operator $ikA_{nm}+k^2\mathcal{D}_{nm}$ can be treated as a perturbation to $\Gamma_{nm}$ given by~\eqref{eq:matrix_element_taunm}. In that regime of particular interest are corrections to the three zero modes, which for finite, but small $k$ become the modes of NS hydrodynamics. Indeed, computing corrections up to the second order in $k$ one finds eigenvalues exactly given by~\eqref{eq:NSdisp}, meaning that $\Lambda_0(k) \approx \Lambda^{\rm NS}_{\rm th}(k)$, and $\Lambda_{1,2}(k) \approx \Lambda^{\rm NS}_{\rm -,+}(k)$~\cite{Lebek_PRL_2025}. In the found expressions, the transport coefficients take exactly the form given by ~\eqref{eq:transport_coeff_diff}.

In the opposite limit of relatively large $k$, it is $\Gamma_{nm}$ that becomes a negligible correction to $ikA_{nm}+k^2\mathcal{D}_{nm}$ and thus we recover the GHD spectrum. This simple analysis is the basic intuition behind explanation of the GHD-NS crossover.

An exemplary plot of real part of the eigenvalues is presented in Fig.~\ref{fig:Spectrum_crossover}(a). We clearly see these two distinct regimes. For very small momenta, there are gapless modes (which we identify as the NS normal modes) and the rest is clearly gapped. At larger $k$, we again notice modes with quadratic dispersions, this time given by GHD diffusion matrix. 

Accordingly, we propose here the following criterion for the GHD-NS crossover. The crossover occurs when the NS dispersion relations~\eqref{eq:NSdisp} hit the line set by the RTA gap $\tau^{-1}$.
This leads to the following formula
\begin{equation}
    k_c = {\rm min} \left[(\mathscr{D}_\pm \tau)^{-1/2},(\mathscr{D}_{\rm th} \tau)^{-1/2} \right],
\label{eq:k_critical}
\end{equation}
where $\mathscr{D}_{\rm th} =\frac{\kappa}{\varrho c_P}$ and $\mathscr{D}_{\pm} =\frac{1}{2 \varrho} \left [ \left (\frac{1}{c_V}-\frac{1}{c_P} \right ) \kappa +\zeta  \right]$. In Fig.~\ref{fig:Spectrum_crossover}(a), we see that this way of defining $k_c$ qualitatively reproduces the crossover point found in numerical diagonalization of $\mathcal{M}(k)$. The crossover happens also in the structure of normal modes, which is contained in the matrix $R(k)$. This is shown in Figs.~\ref{fig:Spectrum_crossover}(b)-(d). In particular note the structure in the $k \ll k_c$ regime in Fig.~\ref{fig:Spectrum_crossover}(c), where three conserved modes clearly decouple from higher charges.
\subsection{Dynamics of small perturbations}
Let us now introduce the initial state into the discussion and analyze the main features of the dynamics given by \eqref{eq:charge_dyn}. The vector $R_{mp}(k)\delta \tilde{q}_p(k,0)$, which enters the expression is a decomposition of the initial state in the eigenbasis of $\mathcal{M}(k)$. As we shall see, the emergence of the Navier-Stokes equations is driven by an interplay of two mechanisms. The first one is a {\em homogenisation}, namely the decay in time of large momentum modes with $k > k_c$. This occurs in the dynamics of all the charges. The second mechanism, that we call {\em thermalisation}, is a decay of charges broken by the collision integral and is important for the dynamics of the small momentum modes. To simplify the picture, we will first look at the dynamics of initial states which are localized in the momentum space such that $\{\delta \tilde{q}_p (k,0)\}_{p \in \mathbb{N}} = 0$ for $|k|$ of order $k_c$ and larger.  After that we will look into dynamics of a generic initial state.

\subsubsection{\texorpdfstring{Initial states with $k_{\rm ini}\ll k_c$}{Initial states}}

For such large scale states we are deep on the left of crossover point in Fig.~\ref{fig:Spectrum_crossover}(a) and the spectrum consists of three NS gapless modes, the rest is gapped. 
Crucially, the sum over $p$ can be truncated due to the structure of matrix visible in Fig.~\ref{fig:Spectrum_crossover}(c), namely
\begin{equation}
    R_{\sigma p}(k \ll k_c)\approx 0,
\end{equation}
for  $\sigma = +,-,{\rm th}$ and $p>2$ or  $\sigma>2$ and $p<2$. We also observe that these matrix elements of $R$ matrix coincide with those of $R^{\rm NS}$. The decoupling in $R$ matrix motivates studying the cases of conserved $n=0,1,2$ and non-conserved $n>2$ charge cases separately. For $n=0,1,2$ we have
\begin{equation}\label{eq:charge_dyn_consv}
\begin{aligned}
&\delta \tilde{q}_n(k ,t) \approx \sum_{\sigma,p} (R^{\rm NS})^{-1}_{n\sigma}(k) e^{-\Lambda^{\rm NS}_{\sigma}(k) t}R^{\rm NS}_{\sigma p}(k)\delta \tilde{q}_p(k,0).
\end{aligned}
\end{equation}
Note that~\eqref{eq:charge_dyn_consv} is nothing else but the NS dynamics~\eqref{eq:charge_dyn_NS}.
Therefore, it is clear that we recover NS hydrodynamics in GHD-Boltzmann equation. 

Next, we consider the $n>2$ case. There we find just the exponentially decaying dynamics
\begin{equation}\label{eq:charge_dyn_nonconsv}
\begin{aligned}
\delta \tilde{q}_{n>2}(k ,t) \approx e^{-t/\tau}\delta \tilde{q}_n(k,0).
\end{aligned}
\end{equation}
It is interesting to observe that in the $k_{\rm ini} \ll k_c$ regime the conserved and non-conserved charges decouple and one does not have to be in the regime $t\gg \tau$ in order to see that the conserved fields with $n=0,1,2$ obey NS hydrodynamics. This is demonstrated in the upper row of Fig.~\ref{fig:comparison_fields}. Note however, that only at such times the perturbations of the higher charges has decayed and the state is fully characterized by the NS degrees of freedom.

\subsubsection{Generic initial states}
We now move to the discussion of more general states, which are not necessarily localized in the region $|k| \ll k_c$ and involve higher Fourier components. The components with $k \gg k_c$ decay on timescales given by GHD dispersion relations. Even though such modes are typically thought of as long-lived, we have to be aware that this is the statement for asymptotically small $k$. The timescales related with decay of $k \gg k_c$ GHD modes are still faster than $\tau$, see Fig~\ref{fig:Spectrum_crossover}(a). Thus at $t \gg \tau$ all Fourier components with $k>k_c$ decay. This puts us back into the setting analyzed in previous section and we still get the NS dynamics~\eqref{eq:charge_dyn_consv} for conserved fields with $n=0,1,2$. The modes with $n>2$ obviously decay to 0.

The dynamics~\eqref{eq:charge_dyn_consv} is thus the universal fate of any initial state in the GHD-Boltzmann framework. The NS hydrodynamization, that is the emergence of NS hydrodynamics happens due to the decay of higher charges as well as higher Fourier components. Eventually, at infinite time only $k=0$ component remains and system becomes homogenous, as expected from hydrodynamics. This picture is confirmed numerically in Fig.~\ref{fig:comparison_fields}, where we compare the corresponding fields obtained numerically by evolving the linearized GHD–Boltzmann equation \eqref{eq:GHD_linear} and the linearized NS equations \eqref{eqs:NS_linearized}. All coefficients entering the NS equations are evaluated from the thermodynamics and collisional matrices as given by~\eqref{eq:transport_coefficients}. We find excellent agreement between them for times $t \gg \tau$. On the other hand, for small times, the dynamics is well captured by GHD, see insets in the bottom row of Fig.~\ref{fig:comparison_fields}.

\section{Hydrodynamic correlation functions} \label{sec:correlations}
In this section, we supplement the discussion of the dynamics by considering the problem of hydrodynamic correlation functions~\cite{DeNardis2022}. The central object of interest is the connected correlator of local charges
\begin{equation}\label{eq:So1o2}
    S_{o_1,o_2}(x,t) = \langle o_1(x,t) o_2(0,0) \rangle_c,
\end{equation}
where $o_{1,2}$ are densities of some local charges $O_{1,2}.$ The initial value of correlation is given by $S_{o_1,o_2}(x,0)=\delta(x)\langle f_{o_1} | \mathcal{C} | f_{o_2} \rangle$, see~\eqref{eq:chargeo} for definition of $f_{o_{1,2}}$.
The correlator of local fields can be accessed through the standard reasoning of linear response and in our setup we will build on the solution~\eqref{eq:charge_dyn} in order to get it. For convenience, let us consider the following expectation value
\begin{equation}\label{eq:chargeo}
     o(x) = \int \dd \lambda f_o(\lambda) \rho_{\rm p}(x,\lambda)
\end{equation}
with $f_o(\lambda)$ giving the associated single particle eigenvalue. We assume that this quantity can be represented in the orthonormalized basis meaning that
\begin{equation}
    f_o(\lambda) = \sum_n (h_n|f_o) h_n(\lambda)\equiv\sum_n \alpha_n^o h_n(\lambda),
\end{equation}
then the correlator \eqref{eq:So1o2} can be expressed as
\begin{equation}
    S_{o_1,o_2}(x,t) =\sum_{i,j} \alpha_i^{o_1}\alpha_{j}^{o_2} S_{q_i,q_j}(x,t).
\end{equation}
In what follows, we will assume that the hydrodynamic state is described by space-dependent local GGE with density matrix given by
\begin{equation}
    \rho_{\rm GGE}(t) \propto e^{- \int \dd y \sum_n\beta_n(y,t) q_n(y,t)}.
\end{equation}
For such states, the hydrodynamic correlation function can be computed as~\cite{Doyon2018corr,Doyon_scipost_2020}
\begin{equation}
    S_{q_i,q_j}(x,t) = - \frac{\delta q_i(x,t)}{\delta \beta_j(0,0)}.
\end{equation}
To evaluate this, we can use the expression for charge dynamics given by $\eqref{eq:charge_dyn}$.
Using the definition of susceptibility matrix and the fact the we work in the basis, which is orthonormal with respect to  inner produced induced by it, we get the following
\begin{equation}\label{eq:GHD-B_corr}
    S_{q_i,q_j}(x,t) =\int \dd k \,  e^{ikx} \sum_m R^{-1}_{im}(k) e^{-\Lambda_m(k) t}R_{mj}(k). 
\end{equation}
Correlator $S_{q_i,q_j}(x,t)$ is the simplest to analyze and at the same time in principle gives access to an arbitrary correlator of local fields as given in~\eqref{eq:So1o2}. Let us analyze different regimes of such object in what follows.

\subsection{\texorpdfstring{Regime $t \ll \tau$: generalized hydrodynamics}{Generalized hydrodynamics}}
\begin{figure}[t]
    \centering
    \includegraphics[scale=0.57]{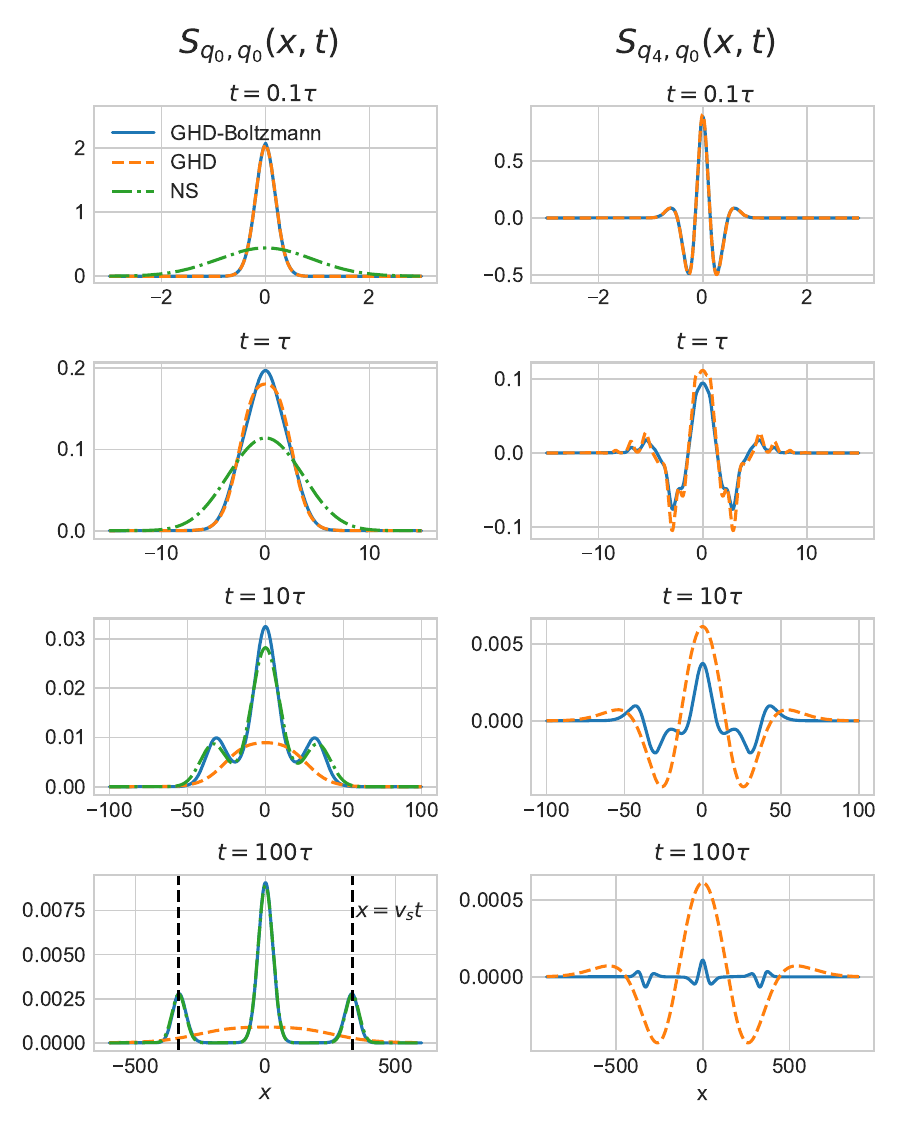}
    \caption{Dynamics of correlation functions of conserved (left column) and non-conserved (right column) charges in GHD-Boltzmann equation. At late times, the density-density correlator exhibits NS behavior~\eqref{eq:S00NS} with two sound modes propagating with sound velocity $v_s$ and heat mode in the center. On the other hand, correlator of non-conserved charge $S_{q_4,q_0}(x,t)$ decays in time. At early times the dynamics in both cases is well described by the GHD without the collision term.}
    \label{fig:correlations}
\end{figure}

For small times $t \ll \tau$, we expect to recover the known GHD results~\cite{Doyon2018corr}. It can be recovered by setting $\Gamma_{nm}=0$ in the $\mathcal{M}_{nm}(k)$ matrix. However, we will use a more convenient formula for GHD correlator at the Euler scale, which reads
\begin{equation}\label{eq:GHD_corr}
    S_{q_i,q_j}^{\rm GHD,E}(x,t) = \int \dd \lambda \delta[x-v(\lambda) t] \rho_{\rm p}(\lambda) f(\lambda) h_{i}^{\rm dr}(\lambda) h_j^{\rm dr}(\lambda).
\end{equation}
In principle, one can compute also the GHD correlation function on diffusive scale. However, it provides only a subleading corrections while introducing technical problems related to computing the exponential of diffusion operator $\mathfrak{D}$. In Fig.~\ref{fig:correlations} we see that indeed in the small time regime the dynamics given by GHD-Boltzmann equation~\eqref{eq:GHD-B_corr} is correctly described by correlations as predicted by GHD. At such short timescales, effects of integrability breaking are still not visible, therefore there are no qualitative differences between correlators involving conserved (non-conserved) charges. The difference is visible at late times, to which we now turn.

\subsection{\texorpdfstring{Regime $t \gg \tau$: conventional fluctuating hydrodynamics}{Regime: conventional fluctuating hydrodynamics}}
At late times the integral in~\eqref{eq:GHD-B_corr} is dominated by contributions from small $k$ modes. Thus the $R$ matrix structure makes all the correlators involving non-conserved charge decay exponentially with the rate set by $\tau$. This is demonstrated in the second column of Fig.~\ref{fig:correlations}. On the other hand, in the case where both charges belong to conserved subspace, the correlator features long-lived behavior. In fact, it approaches the well-known correlator of conventional fluid predicted by fluctuating hydrodynamics. There the hydrodynamic fields are random Gaussian variables with mean fixed by the deterministic dynamics and the variance controlled by the fluctuation-dissipation relation. Since dissipative processes are controlled by the transport coefficients, they set also the strength of the fluctuations. Thus, the transport coefficient are required to actually set up the fluctuating hydrodynamics. For the full exposition see ~\cite{Spohn2014,Mendl2014} or~\cite{chapmanenskog}. Here we write for reference the (normalized) density-density correlator
\begin{equation}\label{eq:S00NS}
\begin{aligned}
    S_{q_0,q_0}^{\rm NS}&(x,t)= \frac{c_P-c_V}{c_P} \frac{1}{\sqrt{4 \pi \mathscr{D}_{\rm th} t}} e^{-\frac{x^2}{4 \mathscr{D}_{\rm th} t}}+ \\
    &\frac{c_V}{2c_P}  \frac{1}{\sqrt{4 \pi \mathscr{D}_\pm t}}\left( e^{-\frac{(x-v_st)^2}{4\mathscr{D}_\pm t}}+ e^\frac{{-(x+v_st)^2}}{{4\mathscr{D}_\pm t}} \right).
\end{aligned}
\end{equation}
Obviously a similar expressions can be obtained for $S_{q_i,q_j}^{\rm NS}$ for $i,j=0,1,2$. In Fig.~\ref{fig:correlations} we show that indeed $S_{q_0,q_0}(x,t)$ becomes $S_{q_0,q_0}^{\rm NS}(x,t)$ at times $t \gg \tau$, in the agreement with the general picture that NS hydrodynamics should be recovered at late times.

As discussed in~\cite{chapmanenskog}, the diffusive form of correlators, such as~\eqref{eq:S00NS}, breaks down at the largest scales. This is the prediction of nonlinear fluctuating hydrodynamics, which incorporates nonlinear terms in the currents of NS equations and emphasizes their impact on correlation functions, which becomes dominant on large scales. In the present context, the nonlinearities become relevant at times of order $O(\tau^2/\tau_{\rm mic})$, where $\tau_{\rm mic}$ is characteristic microscopic timescale of integrable system. Since GHD Boltzmann equation implicitly assumes $\tau \gg \tau_{\rm mic}$ in order for quasiparticle description to hold, there is a parametrically large window, on which linear fluctuating hydrodynamics is expected to provide the correct description.

\section{Conclusions}
In this work, we have addressed two aspects of hydrodynamics under weak integrability breaking. Firstly, we have presented a computation of transport coefficients: viscosity and thermal conductivity in relaxation time approximation for collision integral. Such approximation is motivated by the typically complicated structure of collision integrals stemming from FGR or gBBGKY. The complexity of such collision terms is due to the fact that three particle collisions are the leading thermalization processes in 1D Galilean systems, as opposed to two-particle scatterings in higher dimensions. Moreover, when the collision integral is not known, RTA can still be useful with timescale $\tau$ treated as a phenomenological parameter. The transport coefficients obtained within the RTA can be expressed entirely in terms of the TBA thermodynamics and the Drude weights of the underlying integrable model.

Secondly, we took a close look at the crossover between GHD of infinite conserved charges and NS theory with particle number, momentum and energy being the only conserved quantities. The control parameters of this crossover are space and time scales, which we identified in this work as $k_c$ defined in \eqref{eq:k_critical} and characteristic RTA timescale $\tau$. In the case of RTA,  $\tau^{-1}$ equals the gap of $\mathcal{I}_{\rm RTA}$. The explicit solution of dynamics in the linear regime allowed us to understand the behavior of charge densities. Depending on whether the considered charge is conserved by the perturbation or not, we observed a different dynamics. In the late time regime $t \gg \tau$ the conserved charge densities obey NS dynamics and non-conserved quantities decay exponentially, irrespectively of the initial state. We also characterized the influence of initial state on the dynamics and described the process of homogenisation understood here as a decay of the modes, which are not governed by NS hydrodynamics.

Lastly, we looked at hydrodynamic two-point functions of charge densities. For small times, GHD behavior was found, whereas in the late time regime for conserved charges we found results consistent with linear fluctuating hydrodynamics build upon the NS equations. On the other hand, correlators involving non-conserved charges decay exponentially on timescale set by $\tau$. We also commented on the role of current nonlinearities for the correlation functions of conserved quantities. The effect of such terms is expelled to large space-time scales due to the proximity to integrable model assumed in this work.

Generalizing our RTA computations to other sets of conserved charges is a natural direction for future research. This was done already for the case of Galilean-invariant system with only particle number conservation in~\cite{Lisiak2026}. Thanks to its simplicity, the RTA collision integral can be also useful to study dynamics in non-linear Boltzmann equation, possibly with noise. Such studies could be useful to quantify the role of nonlinearities on various observables at the largest space-time scales.

\begin{acknowledgments}

We thank Leonardo Biagetti and Jacopo De Nardis for discussions on the related topics. S.S,~M.{\L}.~and M.P. acknowledge support from the National Science Centre, Poland, under the  OPUS grant 2022/47/B/ST2/03334. M.Ł. acknowledges support from the National Science Centre, Poland, under the PRELUDIUM grant 2025/57/N/ST2/04184.
\end{acknowledgments}

\appendix

\section{\texorpdfstring{Computation of the matrix elements $\Gamma_{nm}$}{Computation of the matrix elements Gamma}}

\label{app:gamma}
In this section, within the relaxation time approximation, we present the computation of the eigenvalues and the corresponding eigenstates of the operator $\Gamma$ appearing in \eqref{eq:gamma_operator} with elements given in \eqref{eq:ADT_matrix}. We first compute the functional derivative of the collision integral [given by \eqref{eq:collision_integral}] with respect to the rapidity distribution,
\begin{equation}
 \frac{\delta \mathcal{I}_{\rm RTA}[\rho_{\rm p}](\lambda)}{\delta \rho_{\rm p}(\mu)} = -\frac{1}{\tau}\left(\delta(\lambda-\mu)-\frac{\delta \rho^{\rm th}_{\rm p}(\lambda)}{\delta \rho_{\rm p }(\mu)}\right).
\end{equation}
To compute the functional derivative of the thermal distribution $\rho^{\rm th}_{\rm p}$, we decompose it through variations of the charges:
\beq
\frac{\delta \rho^{\rm th}_{\rm p}(\lambda)}{\delta \rho_{\rm p} (\mu)} = \sum_{m,n=0}^2 \underbrace{\frac{\delta \rho^{\rm th}_{\rm p}(\lambda)}{\delta \beta_n}}_{- \mathcal{C} \mathscr{h}_n(\lambda)} \underbrace{\frac{\delta \beta_n}{\delta \mathscr{q}_m}}_{-\mathcal{C}^{-1}_{nm}} \underbrace{\frac{\delta \mathscr{q}_m}{\delta \rho_{\rm p }(\mu)}}_{\mathscr{h}_m(\mu)},
\label{eq:derv_3}
\eeq
where the derivatives are known (as indicated by the underbraces), see~\cite{chapmanenskog}. Using this result, \eqref{eq:derv_3}  simplifies 
\begin{equation}
\begin{aligned}
\frac{\delta \mathcal{I}_{\rm RTA}[\rho_{\rm p}](\lambda)}{\delta \rho_{\rm p }(\mu)} & \notag \\
&\hspace*{-1.2cm}=\frac{1}{\tau} \left[ \sum_{m,n=0}^2 (\mathcal{C}\mathscr{h}_n)(\lambda) (\mathcal{C}^{-1})_{nm} \mathscr{h}_m(\mu) - \delta(\lambda - \mu)\right].
\label{eq:coll_funcl_derv}
\end{aligned}
\end{equation}
The orthonormal charges are defined with respect to the hydrodynamic scalar product involving $\mathcal{C}$. This implies 
\begin{equation}
    \sum_{m,n=0}^2 \mathscr{h}_n(\lambda) (\mathcal{C}^{-1})_{nm} \mathscr{h}_m(\mu) = \sum_{m=0}^2 h_m(\lambda) h_m(\mu).
\end{equation}
Thus, in the bra–ket notation the $\Gamma$ operator in \eqref{eq:gamma_operator} can be written as
\begin{equation}
    \Gamma = \frac{1}{\tau} \left(1-\sum_{m=0,1,2} \mathcal{C}|h_m\rangle \langle h_m|  \right)\mathcal{C}.
\end{equation}
It is clearly diagonal in the orthonormal basis with eigenvalue $0$ for $n=0,1,2$ and eigenvalue $\tau^{-1}$ for $n > 2$.

\section{Simplification of transport coefficients} \label{app:transport}
In this section, we show how we get~\eqref{eq:transport_coeff_coll} from~\eqref{zetaI_simplified} and~\eqref{eq:conductivity_expre_2} of the main text. Here, we present the intermediate steps only for $\zeta_{\mathcal{I}}$. The corresponding steps for the $\kappa_{\mathcal{I}}$ proceed in the same manner. 
Note that the sum in \eqref{zetaI_simplified} runs over all basis states except $0,1,2$. It can therefore be expressed as 
\begin{align}
\zeta_\mathcal{I} &= \tau \langle \mathscr{h}_1 | \mathcal{B}    \left (\sum_{n} | h_n \rangle \langle h_n |-\sum_{n=0,1,2} |h_n \rangle \langle h_n | \right ) | \eta_{\zeta} \rangle, ~\notag \\
&= \tau \langle \mathscr{h}_1 | \mathcal{B} \mathcal{C}^{-1}| \eta_{\zeta} \rangle - \tau \sum_{n= 0,1,2} \langle \mathscr{h}_1 | \mathcal{B} | h_n \rangle \langle h_n | \eta_{\zeta} \rangle.
\label{eq:zetaI_simplified1}
\end{align}
In going from the first line to the second line of \eqref{eq:zetaI_simplified1}, we have used the resolution of the identity
\begin{equation}
    1 = \sum_{\beta} |h_\beta \rangle \langle h_\beta | \mathcal{C}.
\end{equation}
Using the relation between the ultralocal charge basis $\mathscr{h}_1$ and the corresponding orthonormal charge basis $h_1$, namely $\mathscr{h}_1(\lambda)=\sqrt{\mathcal{C}_{11}} h_1(\lambda)=\sqrt{T\varrho} h_1(\lambda)$~\cite{Lebek_PRL_2025}, we can rewrite \eqref{eq:zetaI_simplified1} as 
\begin{align}
\zeta_\mathcal{I}&= \tau \sqrt{T\varrho} \left (\langle h_1 | \mathcal{B} \mathcal{C}^{-1} | \eta_{\zeta} \rangle - \sum_{n= 0,1,2} \langle h_1 | \mathcal{B} | h_n \rangle \langle h_n | \eta_{\zeta} \rangle \right ).
\label{eq:zetaI_simplified2}
\end{align}
From the expression of $\eta_{\zeta}$ in \eqref{eq:eta_functions} and using the fact that $\langle h_m | \mathcal{B} | h_n \rangle $ is nonzero only when $m$ and $n$ are of different parities, we simplify \eqref{eq:zetaI_simplified2} to 
\begin{align}
\zeta_\mathcal{I}&=\tau \varrho  \left (\langle h_1 | \mathcal{B} \mathcal{C}^{-1} \mathcal{B}| h_1 \rangle - \sum_{n=0,2} \langle h_1 | \mathcal{B} | h_n \rangle \langle h_n | \mathcal{B} |h_1 \rangle \right ), \notag \\
&=\tau \varrho  \left (\langle h_1 | D| h_1 \rangle -  \langle h_1 | \mathcal{B} | h_0 \rangle ^2 - \langle h_1 | \mathcal{B} | h_2 \rangle ^2\right ), \notag \\
&=\varrho \tau \omega^{\rm coll}_1.
\end{align}
Note, the Drude weight is defined as $D \equiv  \mathcal{B} \mathcal{C}^{-1} \mathcal{B}$ and the  $\omega^{\rm coll}_{1}$ is given by~\eqref{eq:omega1_simplified}.

\section{\texorpdfstring{Limiting values of $\omega^{\rm coll}_1$}{Limiting values of omega-coll}}
\label{sec:comp_limiting_w1}
It is important to emphasize that the Lieb-Liniger model in the $c\to 0$ limit maps to free bosons while in the $c\to \infty$ limit it maps to free fermions. In this section, we explicitly show that $\omega_1^{\rm coll}$ as given in \eqref{eq:omega1_simplified} vanishes both in the $c\to 0$ and $c\to \infty$ limit.  Note that \eqref{eq:omega1_simplified} can also be written as
\begin{align}
\omega^{\rm coll}_{1} &=\tau \left[ D_{11} -  v^2_s  \right],
\label{eq:omega1_simplified_appen}    
\end{align}
since $v^2_s=A^2_{10}+A^2_{21}$ where $v_s$ denotes the speed of sound and can be  expressed in terms of the known matrices:
\beq
v_s=\sqrt{\frac{1}{\mathcal{B}_{10}}\frac{\mathcal{B}^2_{10} \mathcal{C}_{22}+\mathcal{B}^2_{21} \mathcal{C}_{00}-2 \mathcal{B}_{10} \mathcal{B}_{20} \mathcal{C}_{21} }{\mathcal{C}_{22}\mathcal{C}_{00}-\mathcal{C}^2_{20}}}.
\label{eq:sound_velocity}
\eeq
The matrix elements $\mathcal{B}_{mn}$ and $\mathcal{C}_{mn}$ can be found from \eqref{eq:matrix_susceptibility_element0} and \eqref{eq:matrix_susceptibility_element2}, respectively.
The matrix element $D_{11}$ in \eqref{eq:omega1_simplified_appen} denotes the momentum-momentum Drude weight given in \eqref{drude_weights}.
Now, we show that $D_{11}$ is exactly equal to $v^2_s$ both in the $c\to 0$ and $c \to \infty$ limits. We start with the $c\to \infty$ limit.

\subsection{\texorpdfstring{$c \to \infty$ limit}{c to infinity}}

The scattering kernel, $\mathcal{T}$ in \eqref{eq:scattering_kernel} vanishes in the $c \to \infty$ limit. Consequently, any dressed function $g^{\rm dr}(\lambda)$ in \eqref{eq:dressing_operation}, can be replaced by their original counterparts $g(\lambda)$. As a result all the matrix elements $\mathcal{B}_{mn}$ and $D_{mn}$ can be expressed in terms of the charge-susceptibility matrix $\mathcal{C}$:
\beq
\mathcal{B}_{10}=2\mathcal{C}_{20},\; \mathcal{B}_{21}=2\mathcal{C}_{22}~\text{and}~D_{11}=2\frac{\mathcal{C}_{22}}{\mathcal{C}_{20}}.
\label{eq:relations_elements}
\eeq
We compute $v^2_s$ from \eqref{eq:sound_velocity} using the relations in \eqref{eq:relations_elements} 
\begin{align}
v^2_s
=&\frac{1}{2 \mathcal{C}_{20}} \frac{4 \mathcal{C}^2_{20} \mathcal{C}_{22}+4 \mathcal{C}^2_{22} \mathcal{C}_{00}-8\mathcal{C}^2_{20} \mathcal{C}_{22}}{\mathcal{C}_{22}\mathcal{C}_{00}-\mathcal{C}^2_{20}} 
=2\frac{\mathcal{C}_{22}}{\mathcal{C}_{20}}=D_{11}.
\label{eq:sound_velocity_square}
\end{align}
Therefore, the collisional component of the viscosity vanishes exactly in the limit $c \to \infty$. 

\subsection{\texorpdfstring{$c \to 0$ limit}{c to 0 limit}}
In the $c \to 0$ limit, the scattering kernel, $\mathcal{T}(\lambda-\lambda')$ in \eqref{eq:scattering_kernel} approximates the delta function $\delta (\lambda-\lambda')$. As a result, \eqref{eq:dressing_operation} is simplified to
\beq
g^{\rm dr}(\lambda)=g(\lambda)[1-\mathscr{n}(\lambda)]^{-1}.
\eeq
The integral equation for the pseudoenergy $\epsilon(\lambda)$ in \eqref{eq:epsilon_dressed} changes to an algebraic equation:
\beq
\epsilon(\lambda)=\frac{\lambda^2}{2}-\mu- \log(1+e^{-\epsilon(\lambda)}),
\eeq
from which $\epsilon$ can be solved as
\beq
\epsilon(\lambda)=\log\left(e^{\frac{\lambda^2}{2}-\mu}-1\right).
\eeq
The charge susceptibility matrix elements [in \eqref{eq:matrix_susceptibility_element0}] can then be written as
\begin{align}
\mathcal{C}_{mn}=\int \dd \lambda \rho_{\rm p}(\lambda) \mathscr{h}_m(\lambda) \mathscr{h}_n(\lambda)/(1-\mathscr{n}(\lambda)).
\label{eq:charge_suscep_matrix_appen}
\end{align}
The statistical factor $[1-\mathscr{n}(\lambda)]$ can be related to the bosonic occupation function $\mathscr{n}_b(\lambda)$ as 
\begin{align}
1-\mathscr{n}(\lambda)=1-\frac{1}{e^{\epsilon(\lambda)}+1}=\frac{1}{1+\mathscr{n}_b(\lambda)}
\end{align}
with $\mathscr{n}_b(\lambda)=1/(e^{\epsilon}(\lambda)-1)$. Thus the matrix elements $\mathcal{C}_{mn}$ in \eqref{eq:charge_suscep_matrix_appen} changes to
\begin{align}
\mathcal{C}_{mn}=\int \dd \lambda \rho_{\rm p}(\lambda)  [1+\mathscr{n}_b(\lambda)]\mathscr{h}_m(\lambda) \mathscr{h}_n(\lambda).
\label{eq:charge_suscep_matrix_appen2}
\end{align}
As a result, the statistical factor $f(\lambda)$ in ~\eqref{eq:matrix_susceptibility_element0}, \eqref{eq:matrix_susceptibility_element2} and \eqref{drude_weights} is replaced by $[1+\mathscr{n}_b(\lambda)]$ and  all the calculations in the $c \to 0$ limit resemble that of the $c \to \infty $ limit. We recover the relation: $v^2_s=D_{11}$ and the collisional contribution $\omega^{\rm coll}_1$ vanishes in the limit $c \to 0$ as well.

\bibliography{references.bib}

\end{document}